\documentclass[twocolumn,aps,prl,amsmath,showpacs,nofootinbib]{revtex4}

\usepackage[utf8]{inputenc}
\usepackage{graphicx}
\usepackage{bm}                      
\usepackage{mathptmx}                
\usepackage{amssymb}  
\usepackage{dcolumn}                 
\usepackage{color}
\usepackage{comment}

\newcommand{\be}{\beta}

\newcommand{\beq}{\begin{equation}}
\newcommand{\eeq}{\end{equation}}
\newcommand{\ba}{\begin{array}}
\newcommand{\ea}{\end{array}}
\newcommand{\bea}{\begin{eqnarray}}
\newcommand{\eea}{\end{eqnarray}}
\newcommand{\bi}{\begin{itemize}}  
\newcommand{\ei}{\end{itemize}}
\newcommand{\ben}{\begin{enumerate}} 
\newcommand{\een}{\end{enumerate}}
\newcommand{\bc}{\begin{center}}
\newcommand{\ec}{\end{center}}

\newcommand{\ee}[1]{\times 10^{#1}}

\def\bea{\begin{eqnarray}}
\def\eea{\end{eqnarray}}
\def\be{\begin{equation}}
\def\ee{\end{equation}}

\def\beq{\begin{equation}}
\def\eeq{\end{equation}}
\def\bar{\begin{array}[b]}
\def\barc{\begin{array}}
\def\bart{\begin{array}[t]}
\def\ear{\end{array}}

\begin{document}

\title{Impact of Gravity on Vacuum Stability}

\author{Vincenzo Branchina$^{1,2}$,
Emanuele Messina$^{1,2}$, and D. Zappal\`a$^2$}

\affiliation{$^1$ Department of Physics and Astronomy, 
University of Catania, Via Santa Sofia 64, 95123 Catania, Italy}

\affiliation{
$^2$ INFN, Sezione di Catania, Via Santa Sofia 64, 95123 Catania, Italy}

\date{\today} 

\begin{abstract}
In a pioneering paper on the role of gravity on false vacuum decay, 
Coleman and De Luccia showed that a strong gravitational field can 
stabilize the false vacuum, suppressing the formation of true vacuum 
bubbles. This result is obtained for the case when the energy density 
difference between the two vacua is small, the so called thin wall regime, 
but is considered of more 
general validity. Here we show that when this condition does not hold, 
however, {\it a strong gravitational field (Planckian physics) 
does not necessarily induce a total suppression of true vacuum bubble 
nucleation}.
Contrary to common expectations then, gravitational physics at 
the Planck scale {\it does not stabilize the false vacuum}. These
results are of crucial importance for the stability analysis of the
electroweak vacuum and for searches of new physics beyond the Standard 
Model.

\end{abstract}


\maketitle
{\it Introduction} --
The search for new physics beyond the Standard Model (BSM) is one of the 
most important and challenging questions of present experimental and
theoretical physics. The first run of LHC did not show any sign of new
physics\,\cite{firstrun}. 
The second run, reporting local excess of signal in the diphoton 
channel, has generated a certain expectation and enthusiasm in the
community\,\cite{secondrun}.  
Whether or not this signal will be confirmed by future work,
clues are also needed from the theoretical side. In this respect, crucial 
for our understanding and construction of BSM 
theories is the stability analysis of the electroweak (EW)
vacuum\,\cite{cab,sher,jones,sher2,quiro,alta,isido}. 
The discovery 
of the Higgs boson boosted new interest on this question, making it 
one of the hottest topic in theoretical particle 
physics\,\cite{NNLO,isiuno,isidue,bu,our1,our2,our3,our4}. 
At first, it seemed that the knowledge of the Higgs and top masses,
$M_H$ and $M_t$, was sufficient to determine the stability 
condition of the SM vacuum, that is to determine whether it is a 
metastable state ({\it false vacuum}) or a stable one ({\it true vacuum}). 
It was later realized, however, that this knowledge 
is not enough, as the stability condition of the vacuum is very 
sensitive to unknown new physics interactions at high (or very high, 
say Planckian) energy scales\,\cite{our1,our2,our3,our4}. 

Following a pioneering work of Bender and collaborators\,\cite{bender},
Coleman and Callan studied the decay of the false vacuum in a flat 
space-time background\,\cite{coleman}. 
Later, Coleman and De Luccia extended this analysis to include the impact 
of gravity\,\cite{cdl}. 
The physical mechanism that triggers the false vacuum decay is 
quite simple: due to quantum fluctuations, there is a finite probability 
that a bubble of true vacuum materializes in the sea of false vacuum.

Coleman and collaborators worked within the so called 
``thin wall'' approximation, that applies when the two minima 
of the potential, $\phi_{false}$ and $\phi_{true}$, are such that the 
energy density difference $V(\phi_{false}) - V(\phi_{true})$ is much 
smaller than the height of the potential barrier 
$V(\phi_{top}) - V(\phi_{false})$, where $V(\phi_{top})$ is the 
maximum of the potential between the two minima at $\phi_{false}$ 
and $\phi_{true}$.
They found that the probability of materialization 
of a true vacuum bubble decreases for increasing values of its 
size. In a flat space-time background, this probability turns out  
to be always finite, no matter how large the bubble size. In a 
curved background, things are different.  
In the following we concentrate on the case of a false 
vacuum with vanishing energy density, Minkowski vacuum, and a true 
vacuum with negative energy density, anti-de Sitter (AdS) vacuum, as 
this case is very relevant for applications, in particular for the 
stability analysis of the EW vacuum. 

The transition from a false Minkowski vacuum to a true AdS vacuum 
was studied in\,\cite{cdl}, where it was shown that, when the size of 
the Schwarzschild radius of the bubble of true vacuum is much 
smaller than its size, i.e. when the gravitational effects are 
weak, the probability of materialization of such a bubble is 
close to the flat space-time result, while, when the 
Schwarzschild radius becomes comparable with the bubble size,
i.e. in a strong gravitational regime, the presence of gravity  
stabilizes the false  vacuum, preventing the 
materialization of a true vacuum bubble.
This result was obtained under the thin wall condition mentioned 
above, namely for the case when 
$V(\phi_{false}) - V(\phi_{true}) << V(\phi_{top}) - V(\phi_{false})$. 
However, it is commonly considered as being of more general validity.

Going back to the SM, it is known that due to the top 
loop corrections the Higgs potential $V(\phi)$ bends down for 
values of $\phi >\,\phi_{min}^{(1)} = v $, where $v \sim 246$ GeV 
is the location of the EW minimum, and for the present 
experimental values of $M_H$ and $M_t$, namely $M_H \sim 125.09$ GeV 
and $M_t \sim 173.34$ GeV \,\cite{higgsmass,ATLAS:2014wva}, develops 
a second minimum, much deeper than the EW one and at a much larger 
value of the field, $\phi_{min}^{(2)}\, >>\, v$. 
Clearly, the conditions under which 
the Coleman-De Luccia result is derived are not fulfilled. As said 
above, however, it is still expected that in the presence of 
Planckian physics the bubble nucleation rate 
vanishes\,\cite{espinosa,ridolfi,espinosa2}. In more technical 
words, it is expected that the 
bounce solution to the euclidean equation of motion (from which 
the saddle point approximation for the bubble nucleation rate 
is obtained) disappears.  

Motivated by the above phenomenological considerations, 
in this Letter we focus our attention on the case that is relevant for
the applications to the analysis of the SM vacuum decay, namely the 
decay of a false vacuum with zero energy density to a true vacuum 
with negative energy density, i.e. the decay from a Minkowski false 
vacuum state to an AdS true vacuum. 
To study the impact of gravity on the nucleation rate of a bubble 
of true vacuum, we consider a potential\,\cite{weinberg} 
that allows to explore in a unified framework all possible cases, 
from the thin wall regime to the case when the difference in 
height between the two minima of the potential is large.
For this latter case, we find that the nucleation rate of a true vacuum 
bubble has a value close to the result obtained in flat space-time
{\it even in a strong gravity regime}, when the typical mass scales of 
the potential (see Eq.\,(\ref{potential}) below) become comparable 
with (or even larger than) the Planck scale. 

In fact, we will see that in this case no vanishing of the bubble 
materialization rate is  observed at these high energy scales.
On the contrary, we find that the suppression of the 
Minkowski false vacuum decay (if still present) is pushed 
to very high energy regimes, even much higher than the
Planck scale, so that the possible presence of a Coleman-De Luccia
pole becomes physically irrelevant (similarly to what happens 
for the Landau pole in QED, where the 
latter occurs at such a high energy scale that the theory has 
already lost its significance several orders of magnitudes  
below that scale). At energies below the pole scale, 
but yet Planckian or transPlanckian, the decay 
probability is non-vanishing and very close to the 
flat space-time result, i.e. the result obtained ignoring the 
presence of gravity. 

These findings are new and totally unexpected, 
as it was thought that in a strong gravity regime the decay of a 
false Minkoski vacuum to a true AdS vacuum is always inhibited. 
Moreover, they are of   
crucial importance for current studies and for model building 
of BSM physics, where we are often interested in considering 
new physics at Planckian and/or transPlanckian scales.

\begin{figure}[t]
\vskip 2.5mm
\begin{minipage}{0.4 \textwidth}
\includegraphics[width=.95\textwidth]{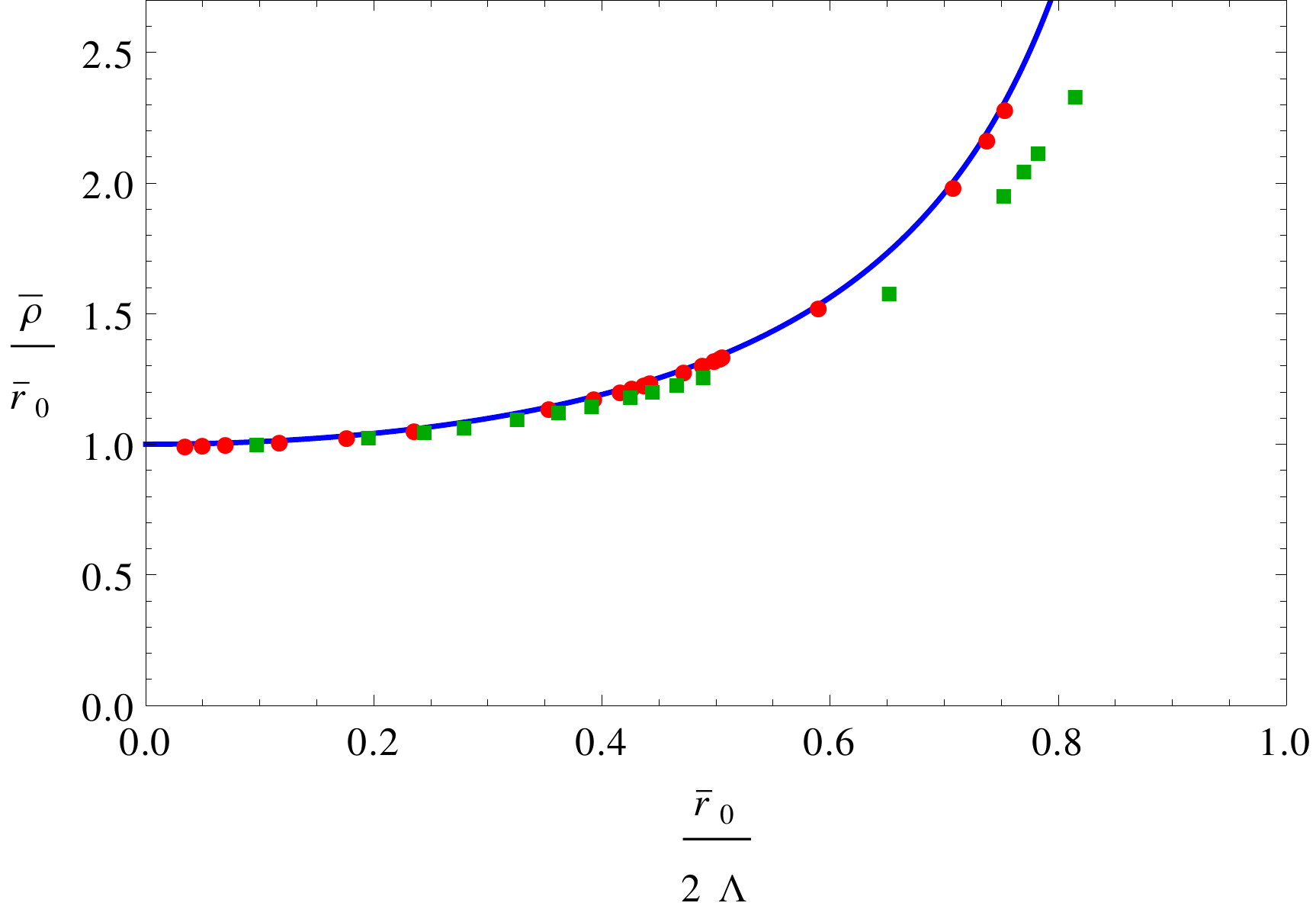}
\end{minipage}
\caption{  $\overline \rho / \overline r_0 $ computed for $b=0.01$ (red online) circles, and for
$b=0.13$ (green online) squares, plotted together with the thin wall prediction, (blue online) solid line.}
\label{fig1}
\end{figure}

{\it Analysis} -- To study the impact of gravity on false vacuum decay, let us consider 
the (Euclidean) action of a scalar field in curved space-time
together with the Einstein term:
\be\label{action}
S=\int d^4 x \,\sqrt{g} \, \left[ \frac{1}{2} g^{\mu\nu} 
\,\partial_\mu \, \phi \; \partial_\nu \phi + V(\phi) 
-\frac{R}{16 \pi G} \right]\,,
\ee
where $R$ is the Ricci scalar, $G$ the Newton constant, and 
$V(\phi)$ the potential\,\cite{weinberg}:
\be
\label{potential}
V(\phi)=\frac{g^2}{4}\left[ (\phi^2-a^2)^2 +
\frac{4\,b}{3}\,(a\,\phi^3 - 3\,a^3\,\phi - 2\,a^4)\right] .
\ee
When $0<b<1$,  $V(\phi)$  has two non-degenerate minima at $\phi=\pm a$, 
$\phi= a$ being the absolute minimum, and a potential barrier that
separates the two minima, with maximum at $\phi=-b\,a$. Note that 
for $b=0$ this 
is just the double well potential with two degenerate minima at 
$\phi=\pm a$, while for $b=1$  the minimum at $\phi=-a$ becomes 
an inflection point and the barrier disappears.    

This potential is perfectly suited for our scopes. By varying the 
dimensionless parameter $b$, we control the difference in height, 
$V(-a) - V(a)$, between the false and the true vacuum. The comparison 
between this difference and the height of the potential barrier, 
$V(-b\,a) - V(-a)$,   
allows to determine whether or not we are in the thin wall regime,  
the latter occuring when $V(-a) - V(a) << V(-b\,a) - V(-a)$.
At the same time, by varying the dimensionful 
parameter $a$ (the only mass scale of our potential), and pushing it 
toward the Planck scale and beyond, we will be able to test the impact 
of strong gravitational physics on the decay of the false vacuum. 

Anticipating from the following analysis, we will see that  
in the $b\to 1$ 
limit (that is the limit of the largest possible difference 
$V(-a) - V(a)$ for the potential (\ref{potential})), the decay 
probability from the Minkowski false vacuum to the true vacuum 
gives exactly the flat space-time result (as if gravity 
was absent!), for {\it any} value of $a$, i.e. from very weak 
to very strong gravitational regimes. For values of $b$ close 
to (but smaller than) $1$, 
the suppression of the vacuum decay should occur for  
very large transplanckian values of $a$. 

\begin{figure}[t]
 \begin{minipage}{0.4\textwidth}
\includegraphics[width=.95\textwidth]{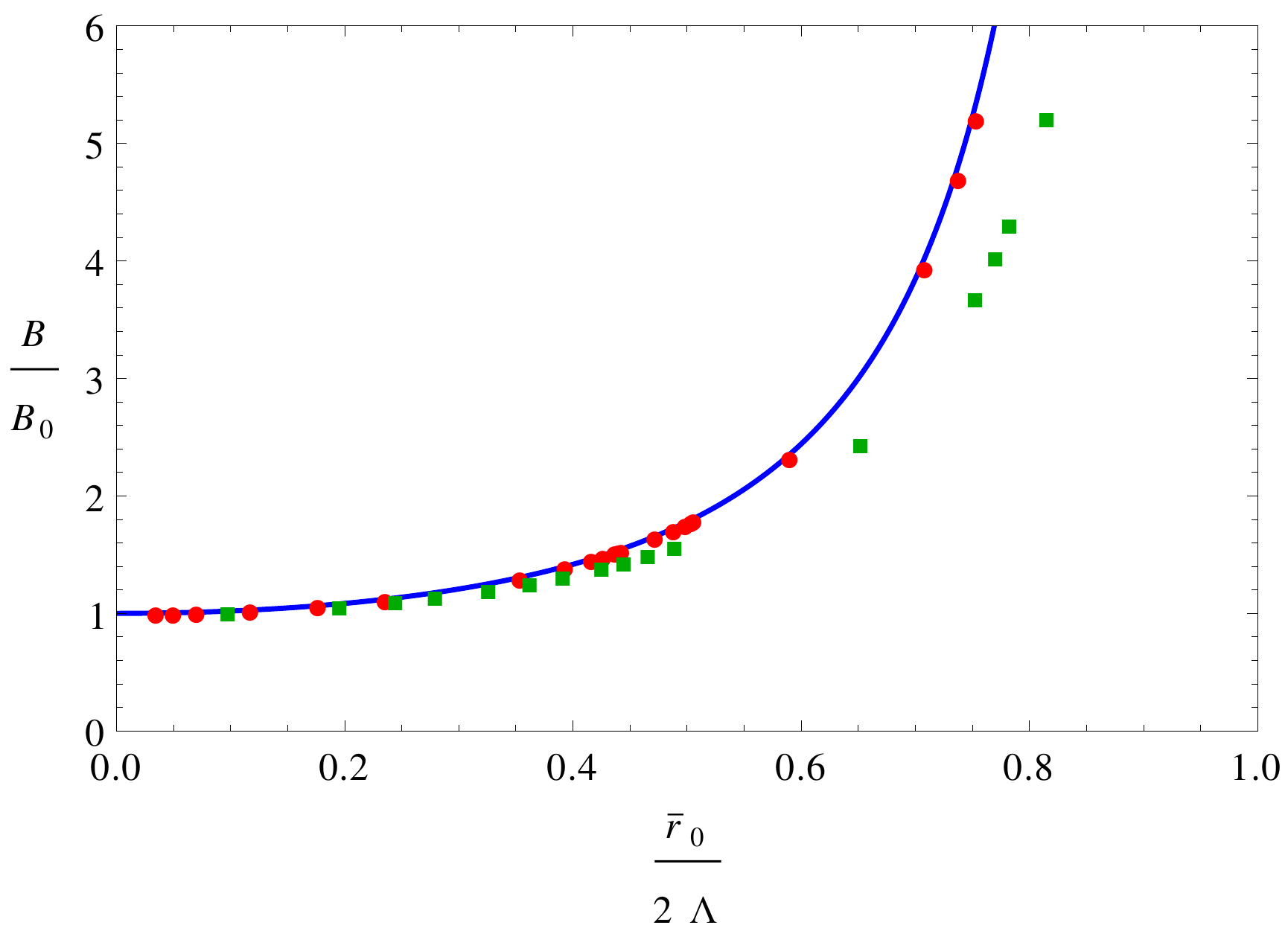}
\end{minipage}
\caption{ $B/B_0$ computed for the same values of the parameters 
used in Fig.\,\ref{fig1}. }
\label{fig2}
\end{figure}

To calculate the  bubble nucleation 
rate $\Gamma$ we need the bounce solution to the (Euclidean) equations of 
motion derived by varying the action in Eq.\,(\ref{action}). 
The saddle point approximation to the transition rate $\Gamma$ 
per unit volume is given by the ratio between the exponential 
of minus the (Euclidean) action evaluated at the bounce and the action
calculated for the false vacuum solution\,\cite{cdl}:    
\be
\label{probabi}
\frac{\Gamma}{\cal V}\,\sim\,e^{-(S_{bounce} - S_{false})}\equiv
\,e^{-B} \; .
\ee

Following\,\cite{cdl}, we consider the most general $O(4)$-symmetric 
Euclidean metric ($r$ is the radial coordinate along a radial curve):
\be\label{metrica}
(ds)^2= (dr)^2 + \rho(r)^2 \,(d\Omega_3)^2
\ee
where $d\Omega_3$ is the element of distance on a unit three-sphere 
and $\rho(r)$ the radius of curvature of each three-sphere. 

\begin{figure}[t]
\begin{minipage}{0.4\textwidth}
\includegraphics[width=.95\textwidth]{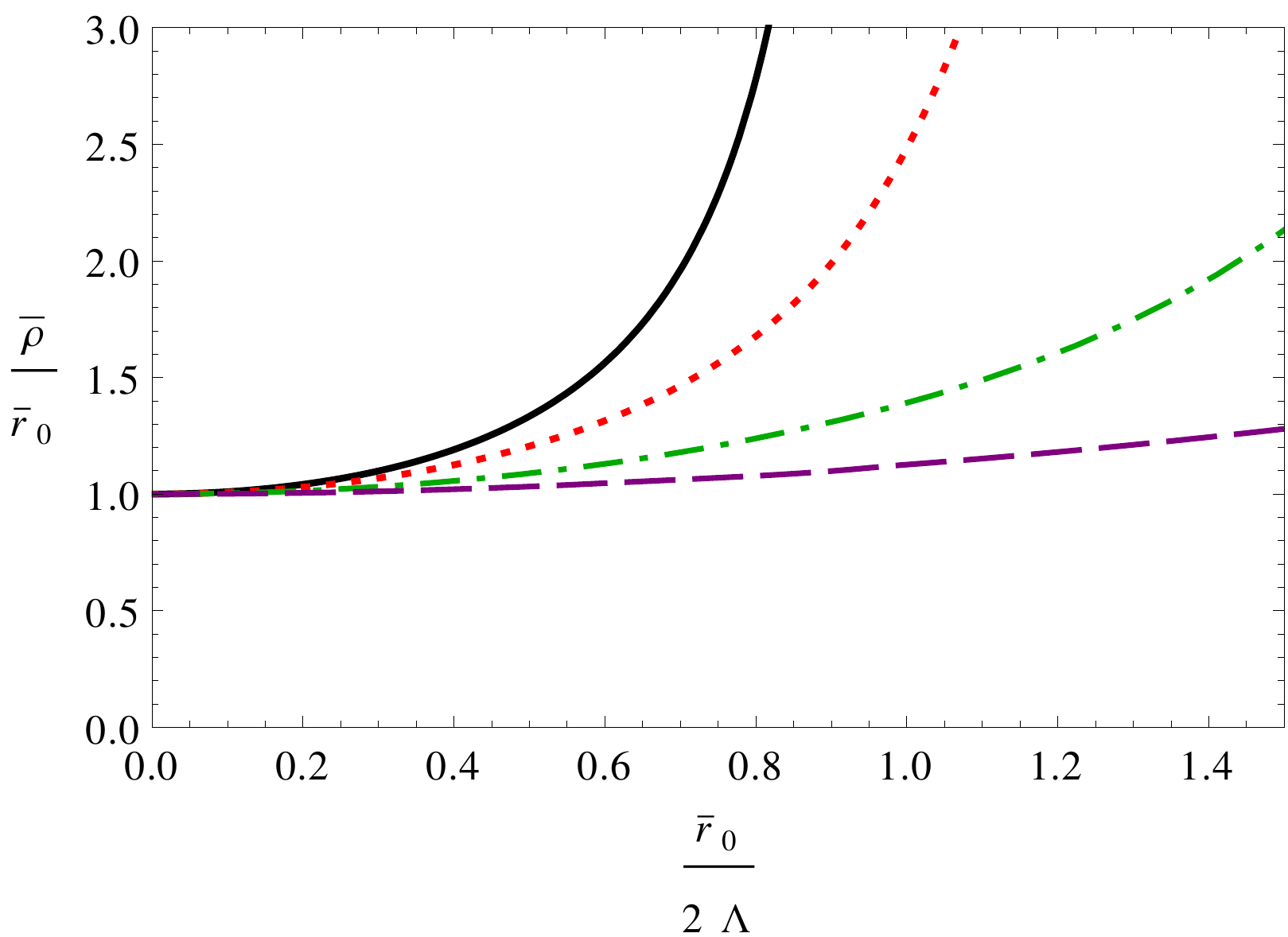}
\end{minipage}
\caption{
$\overline \rho / \overline r_0 $ computed for $b=0.4$ (red online) dotted, 
$b=0.7$ (green online) dot-dashed, 
$b=0.85$ (violet online) dashed line.
The black solid curve corresponds  to the thin wall prediction.}
\label{fig3}
\end{figure}

With the metric (\ref{metrica}), the Einstein equation 
$G_{rr}=-\kappa \, T_{rr}$ reduces to the following equation 
for $\rho(r)$ (the ~$'$ denotes derivation with respect to $r$):
\be
\label{eqmetrica}
{\rho'}^2=1 +\frac{\kappa}{3} \, 
\rho^2 \left (\frac{1}{2} {\phi'}^2 -V(\phi) \right) \,,
\ee
where $\kappa\equiv 8\pi\,G \equiv 8\pi / M_P^2$, 
and $M_P$ is the Planck scale. 
Eq.\,(\ref{eqmetrica}) is coupled to  the scalar field equation 
derived from\,(\ref{action}),
\be
\label{eqmotograv}
\phi'' + \frac{3 \rho'}{\rho} \,\phi' = \frac{dV}{d\phi} \; .
\ee
The false vacuum solution to these coupled equations consists of the
flat space-time metric, $\rho_{false} ' (r) =1$, together with 
the constant solution for the profile, $\phi_{false} (r)= - a$. 

In order to compute $\Gamma$, we have to select the classical solutions of
Eqs.\,(\ref{eqmetrica}) and (\ref{eqmotograv}) that fulfill the boundary 
conditions 
$\phi(+\infty)=-a$,\, 
$\phi ' (0) =0$ \,and\, $\rho(0)=0$.
In our case, the action in Eq.\,(\ref{action}) can be written as:
\bea
S= & \phantom{} & 2\pi^2 \int_0^{\infty} d r  \Bigg[ \rho^3 \, \left ( \frac{1}{2}  \, {\phi' }^2 +V(\phi) \right) +
\nonumber\\
& \phantom{} &\frac{3}{\kappa}  \left ( \rho^2\rho'' +\rho{\rho'}^2 -\rho \right)  \Bigg]=  \nonumber\\
& \phantom{} &       4\pi^2  \int_0^{\infty}    d r \left ( \rho^3 \, V(\phi)  -\frac{3\rho}{\kappa}  \right) + B.T.\,\,,
\label{action_gr}
\eea
where the right hand side is obtained by making use of 
Eqs.\,(\ref{eqmetrica}) and (\ref{eqmotograv}),
and the boundary terms $B.T.$ appear after  integrating  by parts  $\rho^2\rho''$ in the second line 
of Eq. (\ref{action_gr}).

The $B.T.$ are infinite. However, the computation of $\Gamma$ 
only involves the difference $B$ between the 
action evaluated at the bounce and at the false vacuum solutions 
(see Eq.\,(\ref{probabi})). 
These two solutions coincide at infinity, and the two actions provide 
the same divergent contribution, that is then canceled out in the 
difference $B=S_{bounce} - S_{false}$.

It is worth noting that, due to the particular dependence on the 
coupling $g$ of the potential 
$V(\phi)$ in (\ref{potential}), it is 
possible to re-express Eqs.\,(\ref{eqmetrica}) and (\ref{eqmotograv})
in terms of the rescaled quantities  $\hat r = g \,r $,  
$\hat \rho = g \,\rho $ and $\hat V = V /g^2$, so that 
no explicit  dependence on $g$ is left. Accordingly, 
the dependence on $g$ of the action can be 
factored out: $S = \hat S/ g^2$ and $B = \hat B / g^2$.  

\begin{figure}
\begin{minipage}{0.4\textwidth}
\includegraphics[width=.95\textwidth]{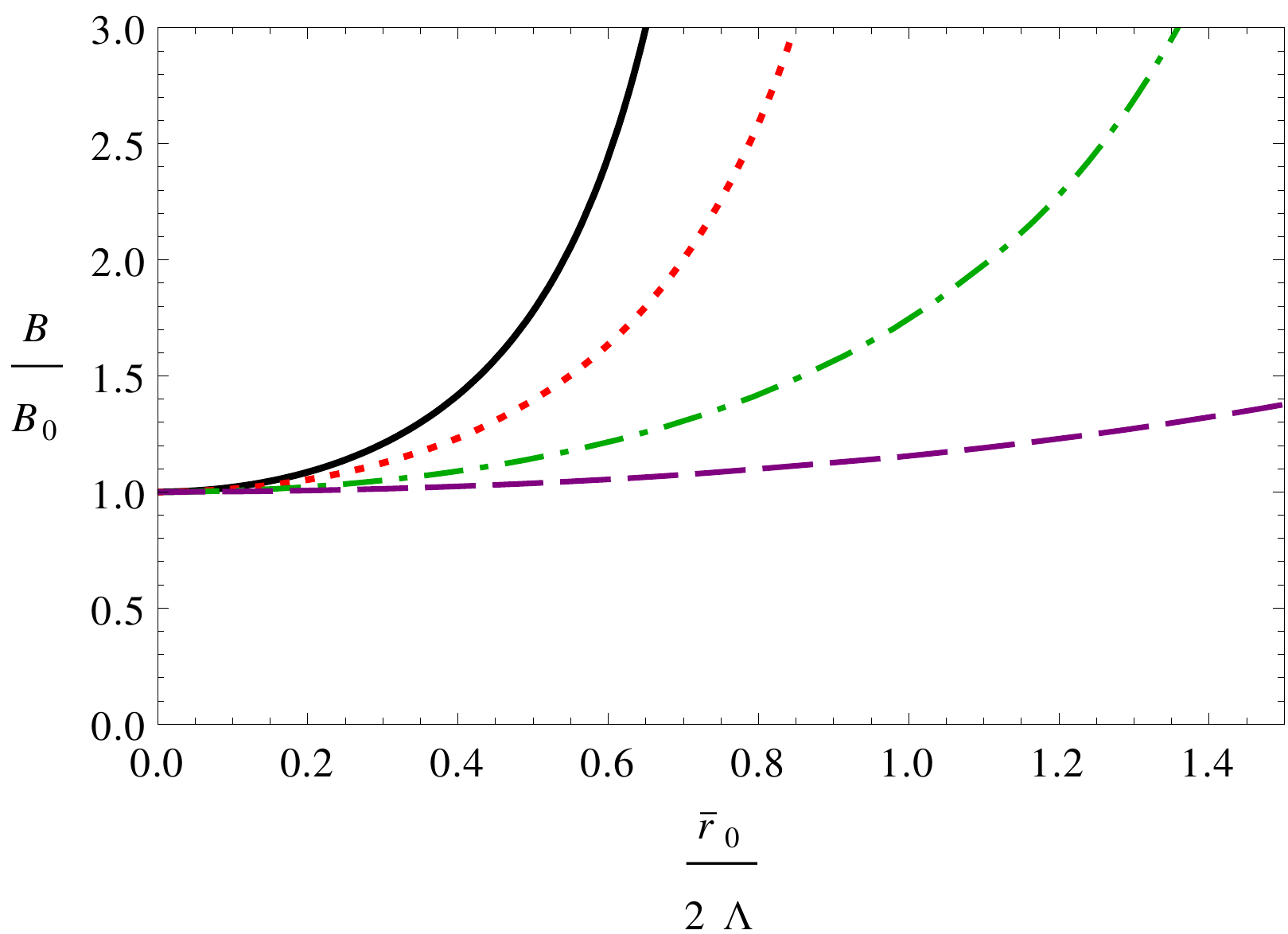}
\end{minipage}
\caption{$B/B_0$ with the same coding as in Fig.\,\ref{fig3}.}
\label{fig4}
\end{figure}

To calculate the transition rate $\Gamma$ for different
potentials of the kind given in Eq.\,(\ref{potential}), i.e.  
for different values of the parameters $a$ and $b$, we need 
first to solve the system of coupled differential 
equations (\ref{eqmetrica}) and (\ref{eqmotograv}) with the boundary 
conditions given above. 

We do that with the help of a numerical shooting procedure.  
The typical bounces are decreasing profiles $\phi_{b}(r)$, 
such that $\phi_{b}(r\to \infty) \to -a$, and can be parametrized by 
their size $\overline r$, the  value of the radial coordinate 
where the height of the profile is decreased of one half.
For large values of $r$, i.e. for $r>> \overline r$, the solution 
for the radius of curvature $\rho(r)$ approaches the flat space-time 
solution, $\rho'(r) =1$, while for  
$r<< \overline r$, i.e. inside the bubble of true vacuum, the presence 
of a negative energy density creates a distortion of this solution.

By defining $\overline \rho$ as the value of the curvature at 
$r=\overline r$, that is $\overline \rho \equiv \rho(\overline r)$, 
a simple way of displaying our findings is to compare the 
results obtained for $B$ and $\overline \rho$ with the corresponding 
quantities derived in the flat space-time case, that will be 
indicated as  $B_0$  and $\overline r_0$ 
 respectively (needless to say, in this latter 
case $\rho (r)$ coincides with $r$).

In the thin wall regime, that for our potential is the case of 
small values of the parameter $b$, it was shown in\,\cite{cdl} that:
\hfill\break 
${\overline \rho}=  {\overline r_0 } \,  \big[ {1 - \left( \overline r_0 / (2 \Lambda) \right)^2}  \big]^{-1} $
\;and \;
$B= B_0  \,  \left [1 - \left( \overline r_0 / (2 \Lambda) \right)^2  \right ]^{-2}$,
i.e.  ${\overline \rho}$ and $B$ have a pole at 
$\overline r_0 / 2 = \Lambda \equiv \sqrt{ 3 / (8 \pi \, G \, |V(a)|}$ .
The corresponding curves are reported in Figs.\,\ref{fig1}, \ref{fig2},
\ref{fig3}, \ref{fig4} as solid lines having an asymptote at 
$ {\overline r_0 }/(2\Lambda)=1$.

Our numerical approach allows to check the accuracy of the 
thin wall predictions, and before moving to the cases 
of particular interest to us, we calculate 
$\overline \rho / \overline r_0$ and $B/B_0$ for some small values 
of $b$ (thin wall regime). 
To facilitate the comparison with\,\cite{cdl}, we present our results 
by plotting $\overline \rho / \overline r_0$ and $B/B_0$ 
against $ {\overline r_0 }/(2\Lambda)$, that is an increasing 
function of $a$ for each fixed value of $b$. The parameter $a$ 
gives the location of 
the minima, and increasing values of $a$ toward the Planck scale 
(and beyond) mark the transition from the weak to the strong 
gravitational regime. 
The results are plotted in Figs.\,\ref{fig1} 
and \ref{fig2}.
Note that the rescaling to the hatted quantities
discussed above shows that  all variables on the $x$ and $y$ axes of 
the figures do not depend on the coupling $g$. 
 
For the smallest value considered for $b$, $b=0.001$, the points 
practically sit on the thin wall curve (blue online solid curve), 
and are not reported in Figs.\,\ref{fig1} and \ref{fig2}. 
At $b=0.01$ (red online circles) we start to observe a small deviation
from the thin wall curve, that becomes
more sizable for $b=0.13$ (green online squares).
Figs.\,\ref{fig1} and \ref{fig2} show that for  
small values of $b$, when the two minima of the potential 
are almost degenerate (thin wall regime), starting  
from $ {\overline r_0 }/(2\Lambda) \sim 0.5$ the effect of gravity
grows rapidly for increasing values of 
$ {\overline r_0 }/(2\Lambda)$. 

We now consider larger values of $b$, thus leaving the thin wall 
regime. Some of the 
curves for $\overline \rho / \overline r_0$ and $B/B_0$ versus 
${\overline r_0 }/(2\Lambda)$ resulting 
from our numerical analysis, namely those obtained for $b=0.4, \,0.7,\, 
0.85$, are reported in Figs.\,\ref{fig3} and \ref{fig4}. These 
figures clearly show that for increasing values of $b$, i.e. 
when the difference in height between the two minima increases, 
both curves $B/B_0$ and $\overline \rho / \overline r_0$ get 
flattened. 

Moreover, for larger and larger values of $b$ ($b\to 1$), 
both curves get closer and closer to the orizontal lines $B/B_0 =1$ and 
$\overline \rho / \overline r_0=1$. We have not added other curves 
in Figs.\,\ref{fig3} and\ref{fig4} as they would unnecessarily clutter 
these figures. For the same reason, we have also limited the 
range of values of ${\overline r_0 }/(2\Lambda)$ to 
$0< {\overline r_0 }/(2\Lambda) < 1.5$. In fact, the Coleman-De Luccia 
pole (of the thin wall case) occurs at 
${\overline r_0 }/(2\Lambda)= 1$, and Figs.\,\ref{fig3} and\,\ref{fig4} 
clearly show 
that this point is not a pole for large values of $b$. 

Finally, for $b\to 1$ these curves reach the horizontal 
lines $B/B_0 =1$ and 
$\overline \rho / \overline r_0=1$ for the whole range of values 
$0 < {\overline r_0 }/(2\Lambda) < \infty$.
This latter result is immediately found by considering the 
bounce profile function $\phi_{b}(r)$, that connects the center 
of the bounce, $\phi_{b}(0)$, with the false vacuum state at 
$r\to \infty$, $\phi_{b}(\infty)= -a$, and seeking for the 
numerical solution of Eq.\,(\ref{eqmotograv}). For small values 
of $b$, $\phi_{b}(0)$ 
is close to $a$, the true vacuum, while for increasing 
values of $b$, $\phi_{b}(0)$ decreases toward 
$\phi_{b}(0) \sim - a$. In the limit $b\to 1$, $\phi_{b}(0)$ reaches 
this latter value, so that 
\be\label{bounceM}
\lim_{b\to 1}\, \phi_{b}(r)=-a
\ee
in the whole range of $r$. In this limit, Eq.\,(\ref{eqmetrica}) 
becomes $\rho'(r)=0$, and we recover the flat space-time result. 

These are the central results of the present Letter. 
In fact, the lesson 
from\,\cite{cdl} (see Figs.\,\ref{fig1} and \ref{fig2}) is that when 
we move from weak to strong gravity regime, the impact of gravity 
becomes enormous, to the point that, for the critical value 
${\overline r_0 }/(2\Lambda)=1$, $B$ diverges and the corresponding 
rate $\Gamma$ for the materialization of a bubble of true vacuum
vanishes. As stressed above, this is expected to hold even out of 
the thin wall regime, and models of BSM physics and conclusions on the 
vacuum stability analysis are often based on this 
expectation\,\cite{espinosa,ridolfi,espinosa2}.

However, we have seen that, when the energy density difference 
between the two vacua of the potential increases (so that the 
conditions of\,\cite{cdl} are no longer verified),
the gravitational contribution to the transition rate 
$\Gamma$ becomes smaller and smaller. Moreover, when this difference 
becomes sufficiently 
large, that for our model occurs for $b \to 1$, the quencing 
of the vacuum decay does not occur even in very strong gravity 
regimes! 

As mentioned before, these results are very surprising and unexpected, 
and also very important for phenomenological applications. In fact, 
extrapolating from\,\cite{cdl}, it was thought that 
whenever we enter in a strong gravity regime, i.e. when we approach 
(and possibly go beyond) the Planck scale, there should always be 
a strong suppression and even a vanishing of the false vacuum  
decay probability, resulting in a stabilization of the false vacuum. 
On the contrary, we see that for potentials where the difference 
in the depth of the minima is sufficiently large, this suppression 
(if still present) is pushed to very high regimes, even (much) above 
the Planck scale, so that the presence of a pole of $B/B_0$ 
becomes physically irrelevant\footnote{In\,\cite{banks} arguments 
are given in favour of the existence 
of such a pole. However, in the cases of interest considered in the 
present paper, this pole would be in a region where the theory is no 
longer valid.}. 

Moreover, the fact that in these conditions the 
transition rate from the false 
to the true vacuum is so close to the flat space-time result is 
of crucial importance for phenomenological applications. In fact,
the case that we have studied is relevant for the stability analysis 
of the electroweak 
vacuum, also in connection with present searches for BSM physics, where
a stability analysis and a clear understanding of the role played by
gravity is greatly needed.

\begin{table}[htb]
\begin{center}

\begin{tabular}{ |c | c | c | }
\hline 
$\textcolor{blue}{ b }$ & $\textcolor{blue}{ g \;\overline r_0\; a }$ &  $\textcolor{blue}{  g^2 \; B_0} $\\
\hline
\hline
$\;\;\;\;\;\; 0.001 \;\;\;\;\;\;$ & $\;\;\;\;\;\;\; 2121.32 \;\;\;\;\;\;$ & $\;\;\;\;\;\; 4.44132\times 10^{10}\;\;\;\;\;\;$ \\
\hline
$0.01$ & $212.113$ & $4.44094\times 10^7$ \\
\hline
$0.13$ & $16.2379$ & $19921.0$ \\
\hline
$0.4$ & $3.93098$ & $276.288$ \\
\hline 
$0.7$ & $2.73633$ & $70.9009$ \\
\hline
$ 0.85$ & $2.51328$ & $22.2602$ \\
\hline
\end{tabular}
\end{center}
\caption{Set of values of 
$g \, \overline r_0 \,a$ and $g^2\, B_0$ computed at different $b$.}
\label{tab1}
\end{table} 

For the reader's convenience, in Table \ref{tab1} some 
values of $g\,\overline r_0 \,a$ and $g^2\, B_0$ 
for different values of $b$ are reported. These are useful 
parameters to reconstruct physical quantities relevat to 
the analysis that we have presented and to better appreciate the 
results reported in the figures (see below).
These quantities are obtained in the flat space-time case, and
are both $g$-independent (as can be seen by resorting to the hatted 
variables introduced before) and $a$-independent (they are 
dimensionless quantities, and in the absence of gravity  
that brings in the dimensionful 
Newton constant $G$, they cannot depend on the only dimensionful 
variable $a$). Therefore, they 
are univocally determined once $b$ is given.

In particular, the values of $g\,\overline r_0 \,a$ in  
Table  \ref{tab1} are useful 
to establish the relation between 
${\overline r_0 }/(2\Lambda)$, that appears in the $x$-axis of all
figures, and the ratio $a/M_P$ (where $M_P$ is the Planck mass) 
for each value of $b$, that is for each single curve in the figures. 
In fact, by recalling the definition of $\Lambda$,
one obtains the following relation ${\overline r_0 }/(2\Lambda) =  
(2 / 3) \sqrt{2 \pi b} \; ( g \, \overline r_0 \,a ) (a / M_P)
$.  Then, for instance,  the point $ {\overline r_0 }/(2\Lambda) =1$
corresponds for $b=0.001$ to $a\sim M_P/100$,  and for $b=0.85$
to $a\sim M_P/4$ (i.e. we are already in a srong gravity regime).

If we now look at the curves in Fig.\,\ref{fig4}, 
these examples clearly illustrate the central results of our work. 
In fact, we see that for increasing 
values of $b$, at fixed values of ${\overline r_0 }/(2\Lambda)$  
(e.g. ${\overline r_0 }/(2\Lambda)=1$),
the suppression of the gravitational effects is larger when the strong 
gravity Planckian regime ($a \to M_P$ ) is approached. Proceeding to 
consider higher values of $b$ (not displayed in the figure), the 
action $B$, and then the decay rate of the false vacuum, stays closer 
and closer to the flat space-time result for larger and larger values 
of $a$ (see the discussion above in relation with Eq.(\ref{bounceM})), 
even for $a >> M_P$!
Finally, with the help of  $g\,\overline r_0 \,a$ and $g^2\, B_0$ 
in Table\,\ref{tab1}, one can derive the values of 
$\overline \rho$  and $B$  from the various curves in the figures.

Before ending this Letter, we have to say some words on previous 
work\,\cite{rych} on the role of gravity on the EW vacuum decay, 
where a perturbative expansion of 
$\phi_b(r)$ and $\rho(r)$ in powers of $\kappa$ around the flat 
space-time solution was attempted.  
Actually, this analytical method seems to have a serious 
drawback, as the boundary conditions (see above), 
essential for the classical solution to be a bounce and in turn for 
$\Gamma$ in Eq.\,(\ref{probabi}) to be the bubble nucleation rate, 
are not respected already at first order in $\kappa$. Therefore,
the output of this analysis cannot be trusted and a fortiori cannot 
be used for comparison.
A detailed study of this issue goes beyond 
the scope of the present paper and will be presented elsewhere. 

{\it Conclusions} -- 
We have studied the decay of a false Minkowski vacuum to a true 
AdS vacuum by taking into account the impact of gravity.

When the energy density difference between the false and the true 
vacuum is large compared to the height of the potential barrier 
(a condition that in the model discussed in the present Letter 
is obtained for $b \to 1$ in the potential of Eq.\,(\ref{potential})),
we find that the effect of gravity on the bubble production rate
practically disappears even in the strong gravity regime, and the 
flat space-time result is recovered.

From a phenomenological point of view, the case that we have 
studied is of particular interest, as it is close to what 
happens in the stability analysis of the electroweak 
vacuum, and several models considered for physics beyond the 
Standard Model require a stability 
analysis where the role played by gravity needs to be carefully 
understood.

Therefore, the new results presented in this work are of the 
greatest importance 
for present studies on the SM stability problem and for searches of 
new physics beyond the Standard Model. They represent a real progress 
in understanding the impact of gravity on the stability of the vacuum. 
In the past, despite attempts to study this question\,\cite{rych}, 
the role of gravity beyond the thin wall case was poorly understood.

\end{document}